\documentclass{aa}  

\usepackage{graphicx}
\usepackage{subcaption}
\usepackage{txfonts}

\usepackage{cuted}

\usepackage{xcolor}

\usepackage[
  colorlinks=true,
  linkcolor=blue,
  citecolor=blue,
  urlcolor=blue
]{hyperref}
\usepackage{orcidlink}
\newcommand{\orcid}[1]{\orcidlink{#1}}

\newcommand{\AREPO}{\texttt{AREPO}\xspace}

\newcommand{\GADGETFOUR}{\texttt{GADGET4}\xspace}

\newcommand{\PKDGRAV}{\texttt{PKDGRAV3}\xspace}
\newcommand{\StePS}{\texttt{StePS}\xspace}
\newcommand{\StePSIC}{\texttt{stepsic2}\xspace}
\newcommand{\StePSHF}{\texttt{StePS\_HF}\xspace}
\newcommand{\CAMB}{\texttt{CAMB}\xspace}
\newcommand{\pysph}{\texttt{pySPHviewer}\xspace}

\newcommand{\lcdm}{$\Lambda$CDM\xspace}

\newcommand{\mpc}{\mathrm{Mpc}}
\newcommand{\gpc}{\mathrm{Gpc}}
\newcommand{\kmsmpc}{\mathrm{km}/\mathrm{s}/\mathrm{Mpc}}
\newcommand{\rad}{\mathrm{RAD}}
\newcommand{\msol}{\mathrm{M_{\odot}}}

\newcommand{\SoneRtwo}{$\mathrm{S}^1\times\mathbb{R}^2$\xspace}
\newcommand{\Tcube}{$\mathrm{T}^3$\xspace}
\newcommand{\Rcube}{$\mathbb{R}^3$\xspace}

\newcommand{\Nbody}{$N$-body\xspace}

\newcommand*{\abs}[1]{\left| #1 \right|}

\renewcommand*\vec[1]{\ensuremath{\boldsymbol{#1}}}

\begin{document}

   \title{Cylindrical cosmological simulations with \textsc{StePS}}

   \author{G. R\'acz\orcid{0000-0003-3906-5699}\thanks{\email{gabor.racz@helsinki.fi}}\inst{1}
          \and
          V. H. Varga\orcid{0009-0001-7889-5188}\inst{1}
          \and
          B. Pál\orcid{0000-0001-9399-390X}\inst{2,3}
          \and
          I. Szapudi\orcid{0000-0003-2274-0301}\inst{4}
          \and
          I. Csabai\orcid{0000-0001-9232-9898}\inst{2}
          \and
          T. Sawala\orcid{0000-0003-2403-5358}\inst{1}
          }

   \institute{Department of Physics, University of Helsinki, Gustaf H\"allstr\"omin katu 2, FI-00014 Helsinki, Finland
         \and
         Department of Physics of Complex Systems, Eötvös Loránd University, Budapest, Hungary
         \and
         Institute for Particle and Nuclear Physics, HUN-REN Wigner Research Centre for Physics, Budapest, Hungary
         \and
         Institute for Astronomy, University of Hawaii, 2680 Woodlawn Drive, Honolulu, HI 96822, USA\\
         }

   \date{Received: February 20, 2026; Accepted: April 30, 2026}

  \abstract
   {The global topology of the Universe can affect long‑range gravitational forces via boundary conditions. Detailed studies of non-trivial topologies require simulations that natively adopt such geometries. Cosmological $N$-body simulations typically evolve matter in a periodic cubic box. While numerically convenient, this imposes a non-trivial three-torus topology that affects long-range gravitational forces, potentially biasing large-scale statistics.}
   {We introduce a compactified simulation framework that is only periodic along a single axis, characterised by an infinite topology with isotropic boundary conditions towards the perpendicular directions, namely, a \SoneRtwo (slab) topology. This new simulation geometry is ideal for simulating systems with cylindrical symmetries such as filaments or certain anisotropic cosmological models.}
   {We compactified the comoving space via an inverse stereographic projection along the radial direction of a periodic cylinder. Then, we evolved the particles based on Newtonian dynamics. A smoothly varying spatial and mass resolution with radius suppresses edge artefacts at the free outer boundary. Our implementation in the \StePS (STEreographically Projected cosmological Simulations) framework uses a direct $\mathcal{O}(N^2)$ force calculation that maps efficiently to GPUs, as well as an Octree $\mathcal{O}(N \log N)$ force calculation for use on large CPU clusters.}
   {The cylindrical domain’s topology enables fully self-consistent simulations to be run in the \SoneRtwo manifold, while mitigating any periodic-image artefacts with respect to targets whose symmetries are mismatched to a cubic box. The main trade-off is a radially varying resolution with distinct systematics and analysis requirements. Finally, we demonstrate the accuracy of the new simulation method via a standard lambda cold dark matter (\lcdm) cosmological simulation.}
  {}


   \maketitle

\section{Introduction}
\label{sec:introduction}

Cosmological \Nbody simulations are fundamental tools in modern cosmology, providing a numerical framework for studying the non-linear evolution of structure from the linear Gaussian early Universe to the highly non-linear cosmic web observed today. They allow us to test cosmological models, interpret large-scale surveys, and explore scenarios beyond the reach of analytic theory. The first attempts to model gravitational clustering in a cosmological context date back to the works of \citet{1970AJ.....75...13P} and \citet{1974ApJ...187..425P}, who studied the formation of galaxy clusters and introduced a statistical framework for halo formation. These early works laid the foundation for the development of numerical simulations that could capture the complexity of non-linear structure formation.

In the 1970s, the first cosmological \Nbody simulations were performed with relatively few particles and often employed free (open) boundary conditions in spherical regions, as in \citet{1976MNRAS.177..717W} and \citet{1979ApJ...228..664A}. 
At that time, the focus was on modelling galaxy clusters, where edge effects were less critical. By the early 1980s, however, periodic boundary conditions became the standard choice. This shift was driven by the need to mimic an infinite, statistically homogeneous Universe and to avoid artificial edge effects, while also enabling efficient particle-mesh (PM) and hybrid TreePM algorithms that rely on Fourier methods \citep{1983MNRAS.204..891K,1995ApJS...98..355X}. The three-torus (\Tcube $= \mathrm{S}^1\times\mathrm{S}^1\times\mathrm{S}^1$) topological manifold is homeomorphic to the Cartesian product of three circles, and it is often described as a periodic box. Since the Lagrangian of a self-gravitating system in \Tcube is invariant to translation, the linear momentum is conserved in such a system. In the late 20th century, the notion that our Universe has this sort of topological manifold was considered a real possibility  \citep{1984PAZh...10..323Z,1990Natur.343..726B,1997Natur.385..139E}, but modern large-volume galaxy surveys and cosmic microwave background (CMB) observations did not find any signs of periodicity \citep{2002MNRAS.336L..13H, 2016A&A...594A..18P} on the observed scales. Periodic boundary conditions are widely used in such codes as \GADGETFOUR \citep{2022ascl.soft04014S}, \AREPO \citep{2020ApJS..248...32W}, and \PKDGRAV \citep{2017ComAC...4....2P} and remain the dominant approach in large-scale cosmological simulations such as Millennium \citep{2005Natur.435..629S}, Illustris \citep{2014MNRAS.445..175G}, and Euclid flagship \citep{2025A&A...697A...5E}.

Despite their practicality, the three-torus topology is not an ideal representation of the Universe with trivial \Rcube topological manifold. While the cubic \Tcube topology preserves statistical homogeneity, it breaks statistical isotropy, since the cube is invariant only under the octahedral group $\mathrm{O_h}$, not the full continuous rotation group $\mathrm{SO}(3)$. According to Noether's theorem, the angular momentum is generally not conserved in a periodic simulation box, since the Lagrangian of a particle system in this topology is not invariant under continuous rotation. This symmetry-breaking also introduces subtle anisotropies in the gravitational field and can bias clustering statistics, especially when the simulation box is not much larger than the scales of interest. In practice, these effects are mitigated by choosing sufficiently large volumes \citep{2021MNRAS.503.5638R}.

To address these limitations of the most used \Tcube topology, alternative topological manifolds must be explored. \citet{2018MNRAS.477.1949R} proposed the STEreographically Projected cosmological Simulations (\StePS) algorithm, which uses a spherically symmetric \Rcube topological manifold with spatial compactification. This approach preserves full rotational invariance and implements a radially varying resolution: high in the simulation centre and gradually decreasing towards the outskirts, reaching the scale of homogeneity to avoid boundary effects. In this space, the full angular momentum vector of a self-gravitating system is conserved. In this paper, we introduce a new hybrid simulation topology that combines the fully periodic and spherical \StePS approaches: the cylindrical \SoneRtwo topological manifold. In this configuration, the simulation domain is periodic along one axis (the cylinder's axis) and open in the radial direction. This design enables a zoom-in strategy similar to the spherical \StePS approach, with high resolution near the central axis and decreasing resolution outward, while retaining periodicity in one dimension. The Lagrangian of a self-gravitating particle system in this space is invariant under translation along the cylinder's axis and rotations in the $\mathbb{R}^2$ plane, which are described by the $\mathrm{SO}(2)$ group. As a consequence of Noether's theorem, the linear momentum along the cylinder's axis and the angular momentum along the same axis are both conserved. Such a topology is particularly well-suited for studying naturally anisotropic environments such as filamentary structures or anisotropic cosmological models with the same symmetries.

This paper is organised as follows: in Sect.~\ref{sec:algorithm}, we describe the base cosmological simulation algorithm. In Sect.~\ref{sec:ICs}, we present the initial condition generation method. Then, in Sect.~\ref{sec:results}, we introduce the first lambda cold dark matter (\lcdm) simulations in cylindrical \SoneRtwo topological manifold. Finally, in Sect.~\ref{sec:summary}, we summarise our results.

\section{Algorithm}
\label{sec:algorithm}

\subsection{Equations of motion}
The Newtonian equations of motion in comoving coordinates of a system composed of $N$ particles can be written as
\begin{equation}
    m_i \ddot{\vec{x}_i} = \sum^N_{\substack{j=1 \\ j\neq i}} \frac{m_i m_j \vec{F}(\vec{x}_i - \vec{x}_j, h_i + h_j)}{a(t)^3} - 2 m_i H(t) \vec{\dot{x}}_i,
    \label{eq:ComovingEquationsOfMotion}
\end{equation}
where $\vec{x}_i$, $m_i$, and $h_i$ are the coordinate vector, mass, and softening length of the particle with $i$ index, $a(t)$ is the scale factor determined by the background cosmological model,
\begin{equation}
    H(t) = \frac{\dot{a}(t)}{a(t)}
\end{equation}
is the Hubble parameter, and $\vec{F}(\vec{x}_i - \vec{x}_j, h_i + h_j)$ is the smoothed gravitational force between particles $i$ and $j$. This force depends on the $\mathcal{F}(r,h)$ force softening and the background topology. In the \StePS code, the softening,
\begin{equation}
        \mathcal{F}(r, h) = \left\{
                \begin{array}{l l}
			\frac{32{r}^{4}}{{h}^{6}}-\frac{38.4{r}^{3}}{{h}^{5}} +\frac{32r}{3{h}^{3}} & \;\text{\small{if $r < \frac{h}{2}$}},\\
                 \ &\ \\
			-\frac{32{r}^{4}}{3{h}^{6}}+\frac{38.4{r}^{3}}{{h}^{5}}-\\-\frac{48{r}^{2}}{{h}^{4}}+\frac{64r}{3\,{h}^{3}}-\frac{1}{15{r}^{2}} & \;\text{\small{if  $\frac{h}{2}<r<h$}},\\
                \ &\ \\
                \frac{1}{{r}^{2}} & \;\text{\small{if $r>h$}},
                \end{array} \right.
\label{eq:Force_spline_kernel}
\end{equation}
 is used \citep{2019A&C....2800303R}, which corresponds to the cubic spline kernel of \cite{1985A&A...149..135M}. The shape of the $\vec{F}(\vec{r},h)$ force function is determined by the background topology.

For a trivial, flat Euclidean \Rcube space, the smoothed gravitational force law can be written as
\begin{equation}
        \vec{F}_{\mathbb{R}^3}(\vec{r}, h) = -G\mathcal{F}(\abs{\vec{r}}, h)\frac{\vec{r}}{\abs{\vec{r}}},
\label{eq:NewtonianForce}
\end{equation}
where $G$ is the gravitational constant. This formula is equivalent to Newton's law of universal gravitation if $r>h$. The standard flat \lcdm cosmological model assumes this isotropic gravity, and it is possible to use this directly in the
\StePS simulation code. In the case of comoving simulations in this topology, the right hand side of Eq.~\ref{eq:ComovingEquationsOfMotion} has to be extended with a simple radial term,
\begin{equation}
    \vec{F}_r(\vec{x}_i) = +m_i\frac{4\pi G}{3}\rho_0\vec{x}_i,
\end{equation}
that prevents the system from collapsing, where $\rho_0$ is the mean matter density in the simulation volume. For the detailed calculations, we refer to \cite{2018MNRAS.477.1949R}.

While the cosmological principle demands isotropic gravity, most cosmological simulations use non-trivial \Tcube topology for numerical convenience. In a periodic cubic box with a linear size, $L_{box}$, multiple periodic images of the particles have to be taken into account with
\begin{equation}
        \vec{F}_{\mathrm{T}^3}(\vec{r}, h) = \sum\limits_{\vec{n}}-G\mathcal{F}(|\vec{r} - \vec{n}L_{box}|, h)\frac{\vec{r}-\vec{n}L_{box}}{|\vec{r}-\vec{n}L_{box}|}
\label{eq:ForcePeriodic},
\end{equation}
summation formula, where $\vec{n} = \{n_1, n_2, n_3\} \in \mathbb{Z}^3$, $|\vec{r}-L_{box}\cdot\vec{n}|<N_{\mathrm{cut}}\cdot L_{box}$, and $N_{\mathrm{cut}}$ is the cut-off distance of the summation in linear box size units. In practice, this formula converges notoriously slowly and the number of periodic images scales with the third power of $N_{\mathrm{cut}}$. To overcome this limitation, the \cite{1921AnP...369..253E}  summation method is used, in which the force is split into rapidly convergent real‑space and reciprocal‑space parts. While a straightforward implementation of Ewald summation provides sufficient accuracy, evaluating these sums requires hundreds or even thousands of floating-point operations for every interaction. This can be dramatically reduced by pre-computing the difference between the direct (nearest periodic image) forces and the fully periodic Ewald forces on a cubic grid \citep{1991ApJS...75..231H}. By constructing a lookup table from these values, one can use interpolation to efficiently add the Ewald correction term to the quasi-periodic force as
\begin{equation}
    \vec{F}_{\mathrm{T}^3} (\vec{r}, h) = \vec{F}_{\mathbb{R}^3}(\vec{r}, h) + \vec{D}_{\mathrm{Ewald}}(\vec{r}),
    \label{eq:EwaldCorrection}
\end{equation}
where $\vec{r}$ is the separation vector between the nearest image in the \Tcube manifold and $\vec{D}_{\mathrm{Ewald}}(\vec{r})$ is the Ewald correction term. In practice, the long-range forces can be calculated with Fast Fourier Transform (FFT) \citep{CooleyTukey1965, 1988csup.book.....H}, which corresponds to $N_{\mathrm{cut}}\to\infty$, but this method is not implemented in \StePS, since this topology is not the main focus of our code. In the \Tcube topological manifold, the linear momentum is conserved due to translational invariance. However, angular momentum is not conserved because the system's Lagrangian lacks symmetry under continuous rotation. This is reflected in the force law: the force vector between two particles depends not only on their relative positions, but also on their orientation relative to the axes of the periodic box. For the detailed description of \Tcube periodic effects in cosmological simulations, we refer to \cite{2021MNRAS.503.5638R}.

\begin{figure}[h!]
    \centering
    \begin{subfigure}[a]{0.495\textwidth}
        \centering
        \includegraphics[width=0.99\textwidth]{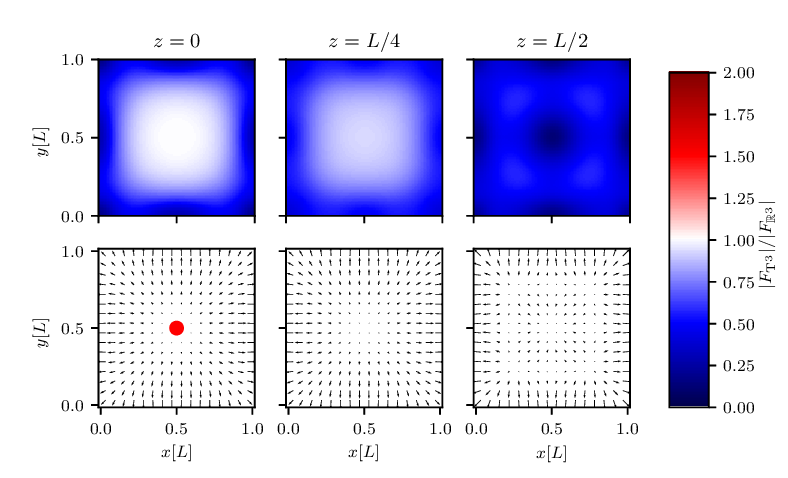}
        \caption{Periodic 3-torus topology}
    \end{subfigure}
    ~
    \centering
    \begin{subfigure}[b]{0.495\textwidth}
        \centering
        \includegraphics[width=0.99\textwidth]{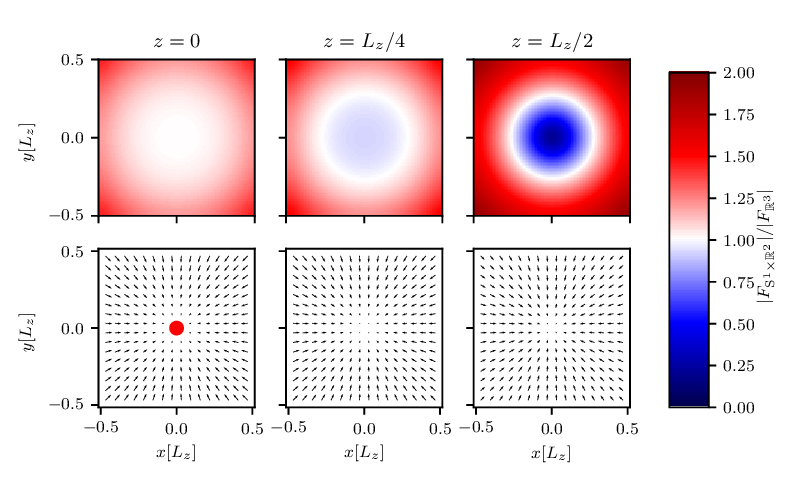}
        \caption{Cylindrical topology, in planes that are perpendicular to the `$z$' symmetry axis.}
    \end{subfigure}
    \centering
    \begin{subfigure}[b]{0.495\textwidth}
        \centering
        \includegraphics[width=0.99\textwidth]{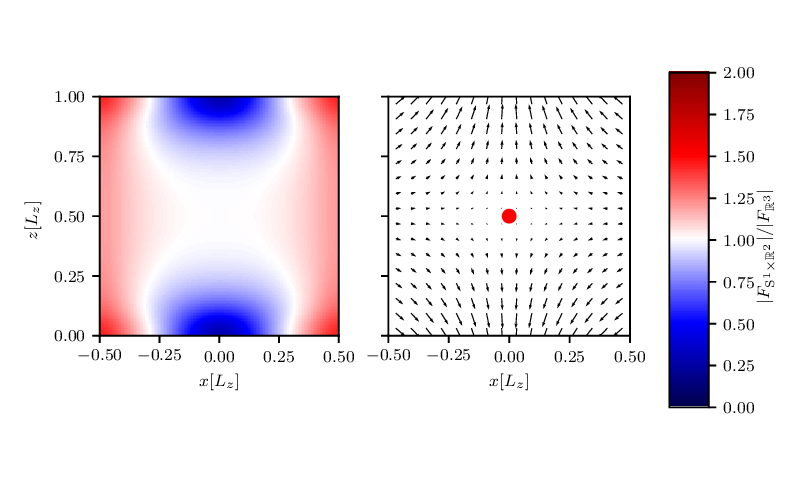}
        \caption{Cylindrical topology, in the $y=0$ plane.}
    \end{subfigure}
    \caption{Comparison of gravitational force laws between the periodic cubic (\Tcube) and the periodic cylindrical (\SoneRtwo) geometry. Each plot demonstrates the deviation from the Newton's law of universal gravitation due to the periodic topology. In both cases, a test particle was placed to the centre of the simulation volume, and the gravitational force field was calculated both in the corresponding and in \Rcube topology. The heat maps are showing the $|\vec{F}|/|\vec{F}_{\mathbb{R}^3}|$ ratio of the force field magnitudes. The vector plots are demonstrating the $\vec{F}-\vec{F}_{\mathbb{R}^3}$ difference of the gravitational force fields around the particle. The length of the arrows is proportional to the magnitude of the difference between the forces.}
    \label{fig:AnisotropicForces}
\end{figure}

In this work, we propose another non-trivial topology for cosmological simulations, the cylindrical \SoneRtwo topological manifold. This space is periodic only along one axis with a periodicity of $L_z$. For simplicity, we chose this as the `$z$'-axis and 
\begin{equation}
        \vec{F}_{\mathrm{S}^1\times\mathbb{R}^2}(\vec{r}, h) = \sum\limits_{\substack{n=-N_{\mathrm{cut}} \\ |r_z-n L_z|< N_{\mathrm{cut}} L_z}}^{N_{\mathrm{cut}}}-G\mathcal{F}(|\vec{r} - n L_z \vec{e}_z|, h)\frac{\vec{r}-n L_z\vec{e}_z}{|\vec{r}-n L_z \vec{e}_z|}
\label{eq:ForceCylindrical}
\end{equation}
summation describes the gravitational force field of a particle in this manifold, where $\vec{e}_z = \{0, 0, 1\}$ is the unit vector in the $z$ direction, $r_z$ is the $z$ component of the $\vec{r}$ vector, and $N_{\mathrm{cut}}$ is the cut-off distance of the summation in linear cylinder size units. While this summation formula still converges slowly, the contribution of distant periodic images now can be directly summed, since the number of images scales only with $N_{cut}$. Nevertheless, the Ewald summation method still can be used in \SoneRtwo for a more accurate force calculation. A comparison between the gravitational forces in $\mathbb{R}^3$, \Tcube, and \SoneRtwo manifolds are shown in Fig.~\ref{fig:AnisotropicForces}. It is important to note that while the Newtonian gravitational forces in a \Tcube space are always weaker than in $\mathbb{R}^3$, the \SoneRtwo forces are weaker only in axial directions, whereas they are always stronger in radial directions.

\begin{figure*}[h!]
    \centering
    \begin{subfigure}[a]{0.32\textwidth}
        \centering
        \includegraphics[width=\linewidth]{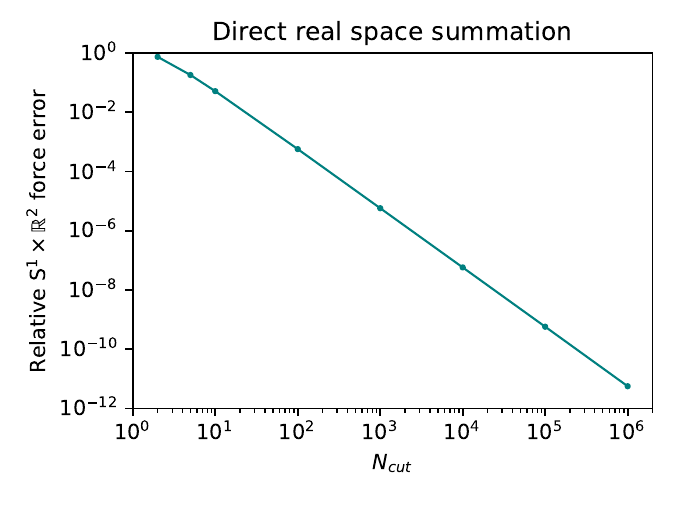}
        \caption{Direct summation of periodic images}
    \end{subfigure}
    ~
    \begin{subfigure}[a]{0.32\textwidth}
        \centering
        \includegraphics[width=\linewidth]{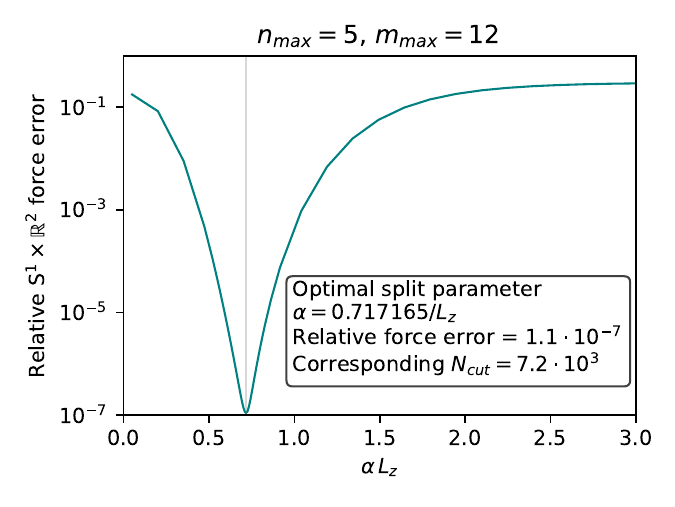}
        \caption{Medium-resolution Ewald summation}
    \end{subfigure}
    ~
    \begin{subfigure}[a]{0.32\textwidth}
        \centering
        \includegraphics[width=\linewidth]{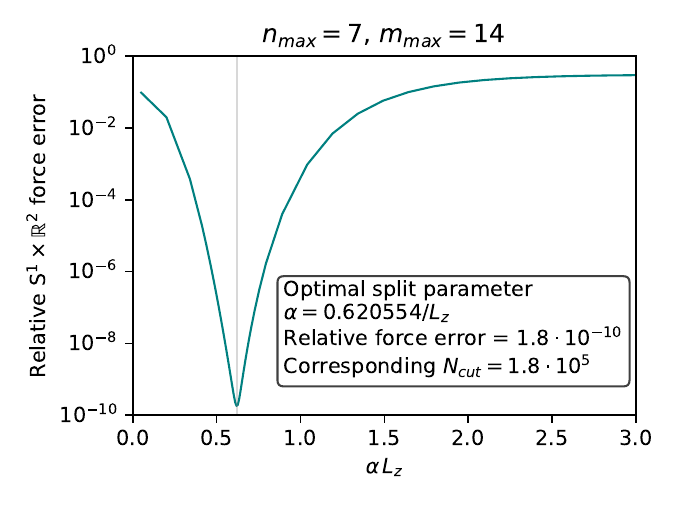}
        \caption{High-resolution Ewald summation}
    \end{subfigure}
    \caption{Accuracy of \SoneRtwo periodic summation methods. For all tests, the correction lookup table $\vec{D}(\rho,z)$ was computed on a $144\times128$ grid, varying the truncation parameter $N_{cut}$ or the Ewald parameters $\alpha$, $n_{max}$, and $m_{max}$. Using these tables, we reconstructed the full periodic force field $\vec{F}_{\mathrm{S}^1\times\mathbb{R}^2}(\rho, z) = \vec{F}_{\mathbb{R}^3}(\rho, z)+\vec{D}(\rho, z)$ and compared it against a reference field. This reference field was generated via direct summation using $N_{cut}=10^7$ periodic images, requiring approximately $10^4\,\mathrm{s}$ of computation time.
    \textit{Left:} Mean force errors of direct periodic real-space summation as a function of $N_{cut}$. Although summation occurs only along a single direction (the $z$-axis), convergence remains exceptionally slow.
    \textit{Center:} \SoneRtwo Ewald summation force accuracy as a function of the Ewald splitting parameter $\alpha$ with fixed $n_{max}=5$ and $m_{max}=12$ values. Using only 5 periodic images in real space and 12 modes in reciprocal space is sufficient to achieve single-precision (32-bit) floating-point accuracy when the optimal splitting parameter is applied.
    \textit{Right:} Same as the center panel, but with fixed $n_{max}=7$ and $m_{max}=14$. Increasing both $n_{max}$ and $m_{max}$ by two improves the force accuracy by nearly three orders of magnitude (a factor of $\sim 600$).
    }
    \label{fig:EwaldAlpha}
\end{figure*}

In comoving cosmological simulations, we only follow the evolution of cosmic structures in comoving coordinates. Since Newtonian gravity is an attractive force, we have to introduce an external force that keeps the simulation from collapsing into the central axis. The net force in every point should be exactly zero for homogeneous matter distribution in comoving coordinates. This can be achieved, if we compensate for the gravitational field, $\vec{F}_{\mathrm{cylinder}}(\vec{r})$, of the homogeneous cylinder in every point. Due to the construction of the summation expressed in Eq.~\ref{eq:ForceCylindrical}, all particles experience the gravitational pull of a cylinder of a height of $2 \cdot N_{\mathrm{cut}}\cdot L_z$, as if they were in the central plane of this cylinder. This gravitational force is calculated by \cite{H_Y_Chang_1988} and takes the form of
\begin{equation}\label{eq:F_horiz}
\begin{split}
    &F_h(\varrho,R,L) = \\ &= 2G\rho_0\left[-2\int^R_0 \ln \left( \frac{\sqrt{L^2+R^2+\varrho^2+2\varrho\sqrt{R^2-Y^2}}+L}{\sqrt{L^2+R^2+\varrho^2-2\varrho\sqrt{R^2-Y^2}}+L}\right) dY \right. \\
    &\left. + \int^R_0 \ln \left( \frac{\sqrt{R^2+\varrho^2+2\varrho\sqrt{R^2-Y^2}}}{\sqrt{R^2+\varrho^2-2\varrho\sqrt{R^2-Y^2}}}\right) dY\right],
        \end{split}    
\end{equation}
for particles within the simulation volume, where $\varrho = \sqrt{x^2+y^2}$ is the radial coordinate measured from the central axis, $L=N_{\mathrm{cut}}\cdot L_{z}$, while $\rho_0$ is the mean matter density and $R$ is the radius of the simulation volume. If $N_{\mathrm{cut}}\to\infty$, then the formula above can be simplified to
\begin{equation}
    F_h(\varrho) = -2\pi G\rho_0\varrho.
    \label{eq:F_horiz_infty}
\end{equation}
Using these formulae, we can write the equations of motion of a self-gravitating \Nbody system in \SoneRtwo topology in comoving coordinates as
\begin{equation}
\begin{split}
    m_i \ddot{\vec{x}_i} =& \sum^N_{\substack{j=1 \\ j\neq i}} \frac{m_i m_j \vec{F}_{\mathrm{S}^1\times\mathbb{R}^2}(\vec{x}_i - \vec{x}_j, h_i + h_j)}{a(t)^3} - 2 m_i H(t) \cdot \vec{\dot{x}}_i \\& - m_iF_h(\varrho, R, N_{\mathrm{cut}}L_z).
    \label{eq:ComovingEquationsOfMotionCylindrical}
\end{split}
\end{equation}
This coupled, second-order differential equation system is solved in our \StePS implementation using a kick-drift-kick (KDK) leapfrog integrator \citep{1983MNRAS.204..891K}, with cosmic time, $t$, as the integration parameter.

\subsection{Ewald summation in \SoneRtwo topological manifold}
To precisely calculate the periodic forces in cylindrical \SoneRtwo topological manifold, one must employ an Ewald summation along the one periodic direction. This problem has been explored in molecular dynamics \citep{2002MolPh.100..635G} and electrostatics \citep{2014arXiv1404.3534T, 2017JCoPh.347..341S}, where these boundary conditions are useful in describing charged particles in strings, pores or channels. In our case, the gravitational force between masses $m_i$ and $m_j$ separated by $\vec{r}$ is split into real- and reciprocal-space parts as
\begin{equation}
    \vec{F}_{\mathrm{S}^1\times\mathbb{R}^2}(\vec{r}) = \vec{F}_{\mathrm{S}^1\times\mathbb{R}^2}^{\rm real}(\vec{r}) + \vec{F}_{\mathrm{S}^1\times\mathbb{R}^2}^{\rm Fourier}(\vec{r}) + \vec{F}_{\mathrm{S}^1\times\mathbb{R}^2}^{m=0}.
    \label{eq:cylindricalEwaldSplit}
\end{equation}
The real‑space part of the Ewald split,
\begin{equation}
    \vec{F}_{\mathrm{S}^1\times\mathbb{R}^2}^{\rm real} (\vec{r}) = -G\,\sum_{n=-n_{max}}^{n_{max}} \frac{\vec{r}_n}{|\vec{r}_n|^{3}}\, \left[ \operatorname{erfc}(\alpha |\vec{r}_n|) + \frac{2\alpha}{\sqrt{\pi}}\,|\vec{r}_n|\,e^{-\alpha^2 |\vec{r}_n|^{2}} \right],
    \label{eq:EwaldRealSum}
\end{equation}
is identical in form to the \Tcube case, but the image set now runs over one integer along the $z$-axis, where $\vec{r}_n = \vec{r}+nL_z\vec{e}_z$ and $\alpha$ is the Ewald-split parameter. In Fourier space, the $z$ component of the wavevector $k_z=\frac{2\pi m}{L_z}$ is discrete due to periodicity in $z$, but continuous in the other components. Carrying out the two-dimensional Fourier integrals, the force contribution at a $\varrho$ radial and $z$ axial separation becomes
\begin{equation}
    \vec{F}_{\mathrm{S}^1\times\mathbb{R}^2}^{\rm Fourier}(\varrho,z) = -\frac{2\,G}{L_z} \sum_{m=1}^{m_{max}} \vec{\nabla} \left[ \cos(k_m z) \mathcal{B}(k_m, \varrho) \right],
    \label{eq:EwaldRecSum}
\end{equation}
where the kernel, 
\begin{equation}
    \mathcal{B} (k_m,\varrho) = e^{k_m \varrho} \text{erfc}\left(\frac{k_m}{2\alpha} + \alpha\varrho\right) + e^{-k_m \varrho} \text{erfc}\left(\frac{k_m}{2\alpha} - \alpha\varrho\right),
    \label{eq:Bkernel}
\end{equation}
avoids the use of modified Bessel functions by utilising the analytical Gaussian filtering instead. The $m=0$ mode represents the background contribution of the periodic mass line, yielding a radial force of
\begin{equation}
    F_{\varrho}^{m=0} = -\frac{2\,G}{L_z \varrho}(1 - e^{-\alpha^2 \varrho^2}).
\end{equation}

While this Ewald summation converges faster than the direct summation of the periodic images, it is still too slow to calculate directly between every particle-particle interaction. To overcome this issue in the \StePS implementation, we build a 2D lookup table,
\begin{equation}
    \vec{D}(\rho, z) = \vec{F}_{\mathrm{S}^1\times\mathbb{R}^2}(\rho, z) - \vec{F}_{\mathbb{R}^3}(\rho, z)
    \label{eq:S1R2EwaldCorrectionTable},
\end{equation}
on an equally spaced grid in $\varrho$ and $z$ cylindrical coordinates, where $\vec{F}_{\mathbb{R}^3}(\varrho, z)$ is the nearest-image Newtonian force. This correction vector is smoothly interpolated during the simulation with cloud-in-cell (CIC) or triangular shaped cloud (TSC) method and added to the nearest image interactions.

Since the direct summation of the periodic images in \SoneRtwo is possible, we used $N_{cut} = 10^7$ real-space periodic images in both direction to construct a brute-force reference lookup-table. We used this to find the optimal $\alpha$ Ewald-split parameters for fixed $n_{max}$ and $m_{max}$ values. This can be seen in Fig.~\ref{fig:EwaldAlpha}.

\subsection{Background evolution}

The background expansion of the Universe is described by the scale factor, $a(t)$, and the Hubble parameter, $H(t)$, and used in Eq.~\ref{eq:ComovingEquationsOfMotionCylindrical}. These can be calculated for any given time by solving the Friedmann equation \citep{1999coph.book.....P, 2008ARA&A..46..385F}
\begin{equation}
\left(\frac{\dot{a}}{a}\right)^2 = H_0^2 \left[ \Omega_{r,0} a^{-4} + \Omega_{m,0} a^{-3} + \Omega_{k,0} a^{-2} + \Omega_{DE}(a) \right],
\label{eq:Friedmann}
\end{equation}
where $\Omega_{r,0}$ is the radiation,$\Omega_{m,0}$ is the non-relativistic matter, $\Omega_{k,0}$ is the curvature, and
\begin{equation}
    \Omega_{DE}(a) = \Omega_{DE,0} \exp \left[ -3 \int_1^a \frac{1 + w_{DE}(a')}{a'} da' \right]
    \label{eq:DEdensity}
\end{equation}
is the dark energy density parameter. In a Chevallier-Polarski-Linder (CPL) parametrisation \citep{2001IJMPD..10..213C, 2003PhRvL..90i1301L}, the $w_{DE}(a)$ dark energy equation of state parameter is written as
\begin{equation}
    w_{DE}(a) = w_0 + w_a (1 - a),
    \label{eq:CPLw0wa}
\end{equation}
where $w_0$ and $w_a$ are free parameters of the model. The $w_a=0$ corresponds to the $w$CDM model, while the cosmological constant (\lcdm) is recovered when $w_0=-1$ and $w_a=0$ in this parametrisation. For the \lcdm, $w$CDM, and general CPL cosmologies, the scale factor, $a(t)$, is determined by solving Eq.~\ref{eq:Friedmann} with a fourth-order Runge-Kutta method in our \StePS implementation. This Friedmann solver uses the same timestep length as the \Nbody integrator. Our current implementation also supports the use of tabulated expansion histories, which can be loaded from an ASCII file. When using these tabulated histories, the code interpolates between the provided values. The interpolation scheme can be configured to use linear, quadratic, or cubic algorithms.

\subsection{Force calculation}

\begin{figure*}[h!]
    \centering
    \begin{subfigure}[a]{0.32\textwidth}
        \centering
        \includegraphics[width=\linewidth]{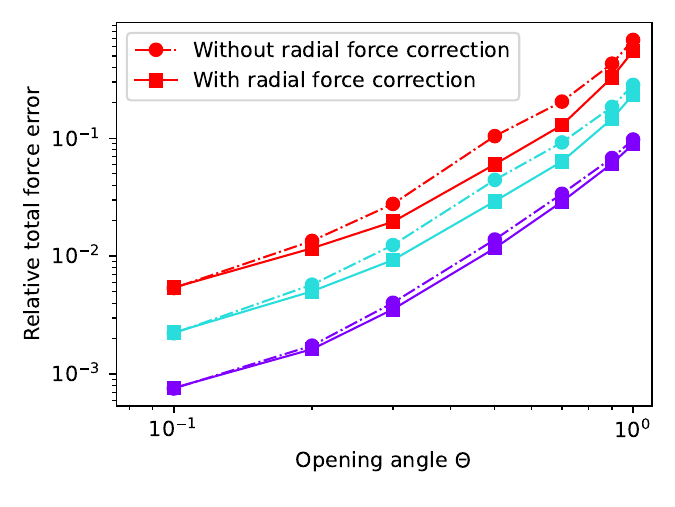}
        \caption{Relative total force error}
    \end{subfigure}
    ~
    \begin{subfigure}[a]{0.32\textwidth}
        \centering
        \includegraphics[width=\linewidth]{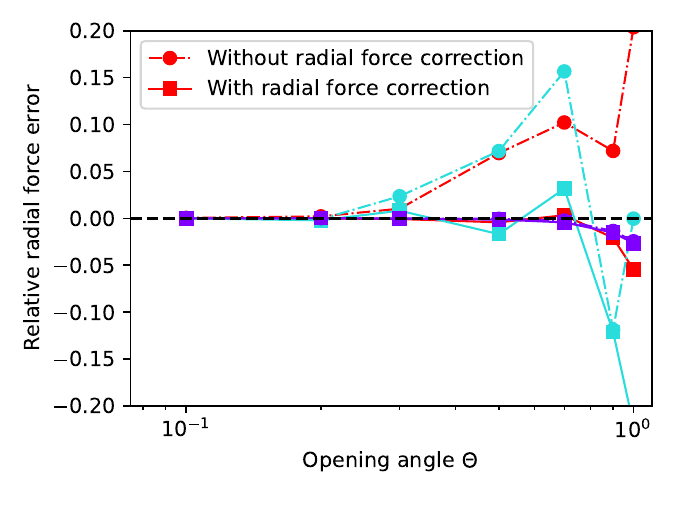}
        \caption{Mean radial force error.}
    \end{subfigure}
    ~
    \begin{subfigure}[a]{0.32\textwidth}
        \centering
        \includegraphics[width=\linewidth]{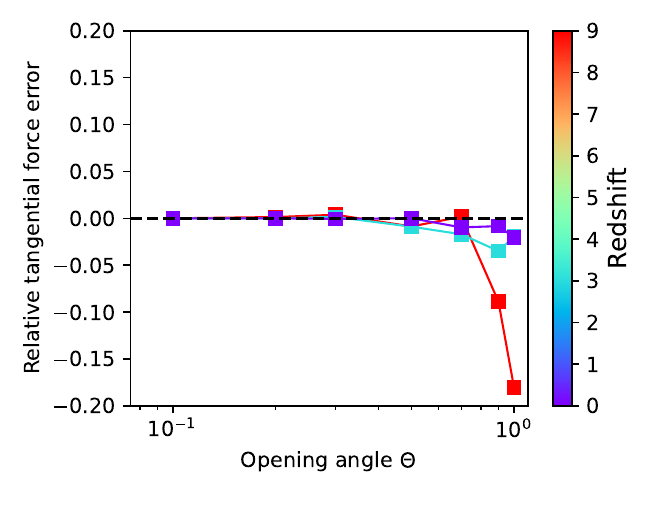}
        \caption{Mean tangential force error.}
    \end{subfigure}
    \caption{Force accuracy as a function of the octree opening angle, $\Theta$, in a cylindrical cosmological volume at different redshifts. We evolved $6\cdot10^4$ particles in standard \lcdm cosmology with direct force calculation in a periodic cylinder with $L_z=200.0\,\mpc$ height and $R_{\mathrm{sim}}=100.0\,\mpc$ radius for this accuracy test. After the simulation was finished, we calculated the $\vec{F}_{BH}$ octree forces with our \StePS simulation code. The $\vec{F}_D$ direct forces were used as a ground truth in the comparison. We calculated the octree forces with and without the radial force correction.
    \textit{Left:} Relative total $\langle|\vec{F}_{BH} - \vec{F}_{D}|/|\vec{F}_{D}|\rangle$ force error. The magnitude of the force error is behaving similarly as in the well tested cubic \Tcube case. The new radial force correction calculated from the initial particle load increases the accuracy in all redshifts.
    \textit{Center:} Mean radial $\langle F_{BH,\varrho} / F_{D,\varrho}\rangle$ force error. While the raw radial octree force have a non-zero expected value at high $\Theta$ values, the implemented radial correction fixes this in all redshifts for $\Theta\leq0.5$. It is critical to use this correction, since a systematic bias in radial direction will cause contraction or expansion of the simulated volume, which will alter the structure formation due to the changed background density.
    \textit{Right:} Mean tangential $\langle F_{BH,\vartheta} / F_{D,\vartheta}\rangle$ force error. In contrast to the radial case, the measured tangential force errors are consistent with a zero mean. A non-zero systematic bias in this component would cause errors in the total angular momentum.
    }
    \label{fig:ForceErrors}
\end{figure*}

To solve Eq.~\ref{eq:ComovingEquationsOfMotionCylindrical}, we have to calculate the acceleration of each particle at every timestep. The calculation of the Hubble drag and the boundary term scales with $\mathcal{O}(N)$, where $N$ is the total number of particles in the simulation. The gravitational interaction between the particles is considerably more complex, since a naive approach requires evaluating all pairwise forces, leading to a computational cost of $\mathcal{O}(N^2)$.

\subsubsection{Direct}
The first public version of the \StePS code supported only direct force calculation, and this option remains available. This was feasible because the original \Rcube \StePS algorithm employs a radially decreasing resolution, allowing a large dynamical range to be covered with only a few million particles. In such a configuration, the computational cost of the $\mathcal{O}(N^2)$ pairwise summation is manageable, especially when combined with efficient GPU parallelization. This approach leverages the zoom-in nature of the method: high resolution is concentrated near the centre, while outer regions are represented by fewer, more massive particles, significantly reducing the total $N$ compared to uniform-resolution simulations.
A direct force calculation was implemented in \StePS for both CPUs and GPUs using MPI-OpenMP and MPI-OpenMP-CUDA parallelisation, respectively. In the newly implemented \SoneRtwo topological manifold, which applies a similar zoom-in strategy in cylindrical coordinates, direct summation remains practical only for smaller cylinders. Because the resolution along the periodic axis is constant, simulations of larger volumes become less efficient with this method. Nevertheless, direct summation is retained in the code as a benchmark due to its simplicity and minimal force-computation errors.

\subsubsection{Octree}

For larger particle numbers, the $\mathcal{O}(N^2)$ force calculation method becomes inefficient even on GPUs. To address this, the new \StePS version implements a \cite{Barnes1986} octree force calculation method for both the original \Rcube and the new cylindrical \SoneRtwo topological manifold to reduce the complexity to $\mathcal{O}(N \log N)$. In our implementation, the octree domains are always cubic and the size of the root node is always twice that of the largest linear dimension of the simulation volume: either the cylinder’s height or diameter in the \SoneRtwo topology, or the simulation diameter in the \Rcube mode.

To minimize systematic errors arising from spatial correlations of the octree geometry, the code introduces random shifts and rotations of the domain at every timestep. In cylindrical topology, the domain center is shifted randomly along the cylinder axis and rotated around the same axis by a random angle. In \Rcube mode, the domain center is shifted randomly within a sphere of a radius $R_{\mathrm{sim}}/2$ from the simulation center, while the octree is rotated around a random axis by a random angle. These transformations help decorrelate force errors and improve the statistical isotropy of the simulation.

In contrast to direct summation, the octree method offers a dramatic improvement in performance for large particle numbers, but at the cost of introducing small force errors that depend on tree parameters and particle distribution. This is particularly important for maintaining radial balance in both the \SoneRtwo and \Rcube cases, where long-range force accuracy governs the stability of the system. While a direct summation automatically satisfies this requirement, the octree-based calculations can lead to spurious radial collapse or expansion if uncorrected. Such a radial instability would change the background density over time during a simulation, which could alter the linear and non-linear power spectrum in a cosmological case. The current \StePS implementation performs a calibration step before the simulation starts: the initial particle load (glass) is used to measure the radial force error under the same octree parameters as the simulation, applying multiple random shifts and rotations. From these measurements, a radial lookup table can be constructed, which is then used during the run to correct octree forces by subtracting the interpolated radial error at every timestep. This correction ensures that the large-scale radial equilibrium is preserved in comoving coordinates without sacrificing the efficiency of the octree. We demonstrate the implemented octree force calculation accuracy with and without a radial correction in a cosmological simulation at different redshifts in Fig.~\ref{fig:ForceErrors}.

The octree implementation is available only with MPI-OpenMP parallelization, making it suitable for CPU clusters. The GPU accelerated octree force calculation will be implemented in the future.

\section{Generating initial conditions}
\label{sec:ICs}

The first step in running a cosmological $N$-body simulation is to generate initial conditions that are consistent with the chosen cosmological parameters. This is typically done in two separate steps: (i) prepare a particle distribution that represents a homogeneous density field and (ii) imprint the density fluctuations predicted by linear theory for the adopted cosmological model. In this section, we outline these steps for the \SoneRtwo topological manifold.

\subsection{Simulation geometry and initial particle load}
\label{sec:PreICs}

\begin{figure}[h!]
    \centering
    \includegraphics[width=1.0\linewidth]{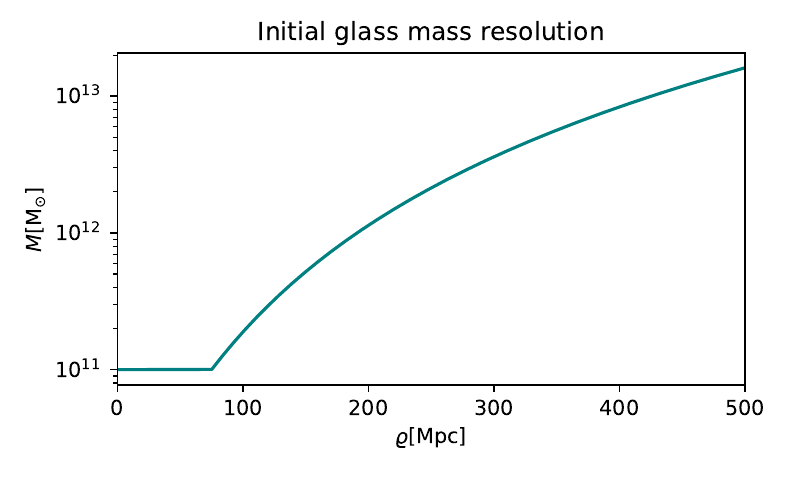}
    \caption{Mass resolution as a function of radial coordinate $\varrho$ in an example cylindrical glass. The parameters of this glass are the following: $D_S=70\,\mpc$, $R_{sim}=500.0\,\mpc$, $\omega_c=1.63973\,\rad$, $L_z = 500\,\mpc$, and the total particle number is $N_{part} = 1.2\cdot10^6$. The glass has constant resolution for $\varrho \, < \, \varrho_c = D_S\tan(\omega_c/2) = 75.0\,\mpc$ to avoid unwanted mixing of particles with widely different masses in the simulations.}
    \label{fig:S1R2StePS_resolution}
\end{figure}

The main simulation challenge in non-finite-volume topological manifolds is to handle the edge effects where the simulation transitions from the unresolved to the resolved volume. The original spherical \StePS algorithm addressed this in \Rcube by compactifying space via the inverse stereographic projection. In \SoneRtwo topological manifold, we applied the same technique in two dimensions to the non-compact $(x,y)$ directions.

The inverse stereographic projection is a bijection between the compact $n$-sphere $\mathrm{S}^n$ and $\mathbb{R}^n$ (excluding the projection point $\mathrm{Q}$). Here, we compactify $\mathbb{R}^2$ to $\mathrm{S}^2$ and work in cylindrical coordinates $(\varrho,\varphi,z)$. Only the radial coordinate $\varrho=\sqrt{x^2+y^2}$ is transformed as
\begin{equation}
    \omega = 2\,\arctan\!\left(\frac{\varrho}{D_s}\right),
    \label{eq:inversestereographic}
\end{equation}
where $D_s$ is the diameter of the $\mathrm{S}^2$ sphere used in the mapping and $\omega \in [0,\pi)$ is the new compact coordinate. The point $\omega\to\pi$ corresponds to $\varrho\to\infty$ and is excluded from the finite computational domain. This corresponds to the last volume element around $\mathrm{Q}$. Next, we tessellate the compactified domain uniformly in the coordinates $(\omega,\varphi,z)$. To avoid excessively steep variations in the non-compact resolution, we enforced a constant mass resolution inside a cylindrical core $\omega < \omega_c$, where $\omega_c$ is the maximum compact coordinate for which constant mass resolution is maintained. We then placed a single particle at a random location in each compact cell and assigned it a mass of $m_i = \bar{\rho}\,V_i$, where $\bar{\rho}$ is the mean comoving density and $V_i$ is the corresponding cell volume in the non-compact $(\varrho,\varphi,z)$ coordinates. This ensures that the compact-to-physical Jacobian of the inverse stereographic map is properly accounted for. Once the sample particles are prepared, we transform the compact radius back to the non-compact radius using a stereographic projection,
\begin{equation}
\varrho = D_s\,\tan{\frac{\omega}{2}}
\label{eq:omegatovarrho}.
\end{equation}
Following this transformation, we have a zoom-in geometry in the non-compact space where the spatial resolution smoothly decreases with the radial $\varrho$ coordinate, until we reach the edge of the resolved region ($R_{sim}$). The distribution of the particle masses for an example realisation as a function of the radial $\varrho$ coordinate can be seen in Fig.~\ref{fig:S1R2StePS_resolution}. Using Eq.~\ref{eq:omegatovarrho}, the non-compact radius of constant resolution $\varrho_c$ can be calculated.

To use this distribution as an initial particle load during the initial condition generation, we required  the net comoving gravitational acceleration on each particle to be approximately zero. The usual method for this is to use this particle distribution as an initial condition for glass-making \citep{1994astro.ph.10043W}. The \StePS simulation code is able to generate such glasses in \SoneRtwo topology by running a comoving cosmological \Nbody simulation with repulsive gravity. During the simulation, the particles are losing energy due to the Hubble-drag term in Eq.~\ref{eq:ComovingEquationsOfMotionCylindrical}, and after $\mathcal{O}(10^5\text{--}10^6)$ expansion factors, they are settling down in a glass-like structure where the net gravitational acceleration of every particle is negligible.

\subsection{Initial perturbations}

Once the initial particle load is ready, we generated the initial perturbations using the \StePSIC code (P\'al et al. in prep.)\footnote{\url{https://github.com/eltevo/stepsic}}. This program supports \Tcube, \Rcube, and \SoneRtwo topologies and can be applied in either the Zel'dovich approximation (1LPT) or second-order Lagrangian perturbation theory (2LPT). Given the cosmology, we can compute the linear matter power spectrum at the time of the initial scale factor ($a_{\rm init}$) as
\begin{equation}
    P_{\rm lin}(k,a_{\rm init}) \;=\; D^2(a_{\rm init})\,T^2(k)\,P_0(k),
    \label{eq:linearpk}
\end{equation}
where $P_0(k)$ is the primordial power spectrum, $T(k)$ is the transfer function, and $D(a)$ is the linear growth function. We then realised a Gaussian random field, $\delta(\mathbf{k})$, following the standard Rayleigh–phase sampling \citep{2005ApJ...634..728S}. For each independent Fourier mode, we draw a random phase $\theta_{\vec{k}}\sim \mathcal{U}(-\pi,\pi)$ and an amplitude, $A_{\vec{k}}$, from a Rayleigh distribution with scale parameter, $\sigma_{\vec{k}}$, chosen so that the real and imaginary parts of $\delta(\vec{k})$ end up independent Gaussian variates with zero mean and variance $\sigma_{\vec{k}}^2 = P_{\rm lin}(|\vec{k}|,a_{\rm init})/2$. We then enforced the Hermitian symmetry $\delta(-\vec{k})=\delta^*(\vec{k})$ to ensure a real configuration-space field and proceeded to compute the displacement and velocity fields.

The initial comoving positions, $\mathbf{x}_i$, and velocities, $\dot{\mathbf{x}}_i$, were obtained by displacing the Lagrangian particle coordinates, $\mathbf{q}_i$, of the initial particle load according to the Zel'dovich expression,
\begin{align}
		\vec{x}_i &= \vec{q}_i + \int \frac{i\vec{k}}{k^2} \delta(\vec{k}) e^{i\vec{q}_i\vec{k}},\\
		\dot{\vec{x}}_i & = \frac{\dot{D}(a_{init})}{D(1)}\int \frac{i\vec{k}}{k^2} \delta(\vec{k}) e^{i\vec{q}_i\vec{k}}, 
	\label{eq:Zeldovich}
\end{align}
or, alternatively, with a 2LPT approximation \citep{1994MNRAS.267..811B, 2010MNRAS.403.1859J}. 

\section{Results}

\label{sec:results}

\begin{figure*}[h!]
    \centering
    \begin{subfigure}[a]{0.95\textwidth}
        \centering
        \includegraphics[width=\linewidth]{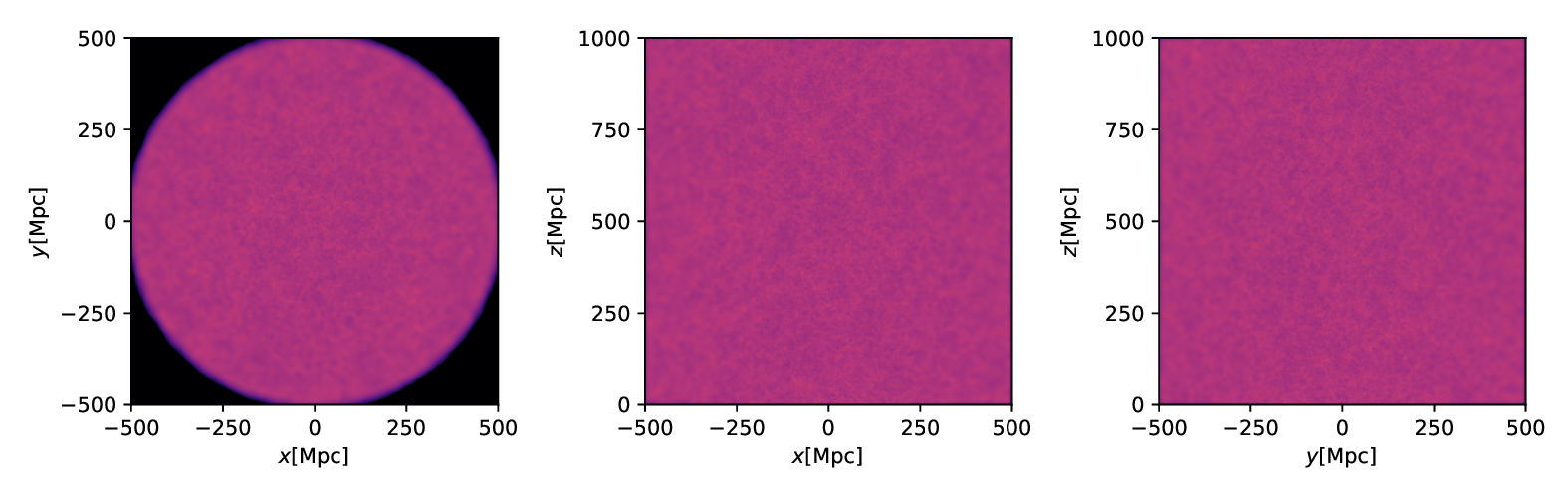}
        \caption{$z=11$}
    \end{subfigure}
    ~
    \begin{subfigure}[a]{0.95\textwidth}
        \centering
        \includegraphics[width=\linewidth]{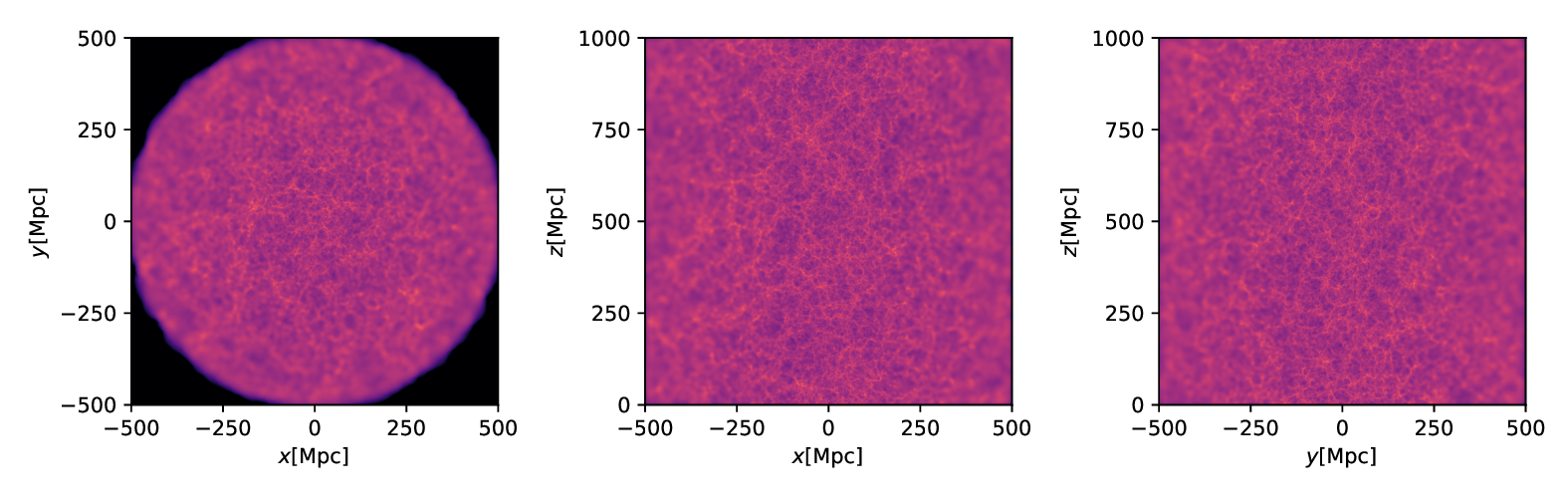}
        \caption{$z=3$}
    \end{subfigure}
    ~
    \begin{subfigure}[a]{0.95\textwidth}
        \centering
        \includegraphics[width=\linewidth]{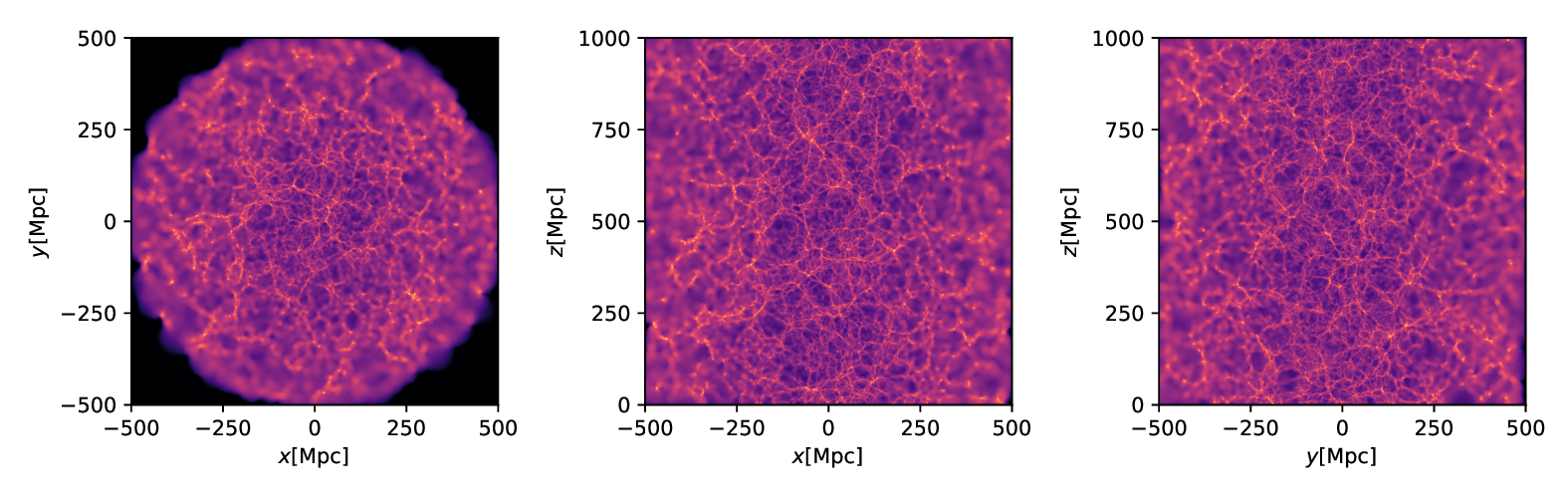}
        \caption{$z=0$}
    \end{subfigure}
    \caption{Thin slices of the dark matter density field from the first \SoneRtwo \StePS simulation at $z=11$, $z=3$, and $z=0$ redshifts in comoving coordinates. Each slice has a thickness of $15.0\,\mpc$ and each column shows a different slice orientation. The late-time non-linear cosmic web is visually similar to that found in \Tcube and \Rcube simulations: voids, walls, filaments, and dark matter halos are clearly visible. Due to the zoom-in nature of the \StePS \SoneRtwo simulation method, the spatial resolution is highest along the central axis, while only the largest structures are resolved near the $R_{\textrm{sim}}$ simulation radius.}
    \label{fig:DMfields}
\end{figure*}

\subsection{\lcdm simulation}

\begin{table*}[h!]
    \caption{Parameters of the first cosmological \Nbody simulation in \SoneRtwo topology.}
    \centering
    \begin{tabular}{| l | l | c |}
         \hline
         \multicolumn{3}{|c|}{Cosmological parameters} \\\hline
         $H_0$ & Hubble constant& $67.66\,\kmsmpc$ \\
         $\Omega_m$ & Matter density parameter & $0.3111$\\
         $\Omega_{\Lambda}$ & Dark energy density parameter & $0.6889$\\
         $\sigma_8$ & Power spectrum normalisation & $0.8102$ \\
         $z_{\mathrm{init}}$ & Initial redshift & 49 \\
         \hline\hline
         \multicolumn{3}{| c |}{Simulation geometry}\\\hline
         $R_{\mathrm{sim}}$ & Simulation radius & $500.0\,\mpc$\\
         $L_z$ & Simulation height & $1000.0\,\mpc$\\
         $D_S$ & Diameter of the compact sphere & $70.0\,\mpc$\\
         $\omega_c$ & Compact radius of constant resolution & $1.63973\,\rad$\\
         $N_p$ & Number of particles & $2.4\,\cdot\,10^7$ \\
         $M_p$ & Particle mass & $1.0\,\cdot\,10^{11} \;-\;1.6\,\cdot\,10^{13}\msol$\\\hline
    \end{tabular}
    \tablefoot{These cosmological parameters were based on the \cite{2020A&A...641A...6P} best-fit cosmological parameters.}
    \label{tab:SimulationParams}
\end{table*}

To validate the new simulation method, we run a new cosmological \Nbody simulation with the updated version of our \StePS code in \SoneRtwo topology. We used the best-fit \textit{Planck} 2018 \citep{2020A&A...641A...6P} \lcdm cosmological parameters and a 2LPT method to generate the initial conditions of a periodic cylinder with a height of $L_{z} = 1.0\,\mathrm{Gpc}$ and resolved radius of $R_{\mathrm{sim}}= 500\,\mpc$. As an initial particle load, we used a cylindrical glass with height $500\,\mpc$ and radius $=500\,\mpc$, repeated along the $z$ axis. The initial power spectrum was calculated by \CAMB (Code for Anisotropies in the Microwave Background) \citep{2000ApJ...538..473L, 2011ascl.soft02026L} at $z=0$, and backscaled to the $z_{\mathrm{init}}=49$ initial redshift with the linear growth function. The parameters of the simulation can be seen in Table~\ref{tab:SimulationParams}. The simulation was run on the \texttt{CSC Puhti}\footnote{\url{https://docs.csc.fi/computing/systems-puhti/}} supercomputer using 640 CPU cores and 3085 timesteps. On average, one integration timestep took $60$ seconds of wall-clock time.
Overall 138 particle snapshots were stored between redshifts 31 and 0. The \pysph \citep{2017ascl.soft12003B} visualisation of the dark matter distribution in this simulation in different slices and redshifts can be seen in Fig.~\ref{fig:DMfields}. In our \SoneRtwo topology, a cosmic web forms from the small initial density fluctuations due to gravity, similarly to the \Tcube and \Rcube simulations.

\subsection{Power spectra}
\begin{figure}
    \centering
    \includegraphics[width=\linewidth]{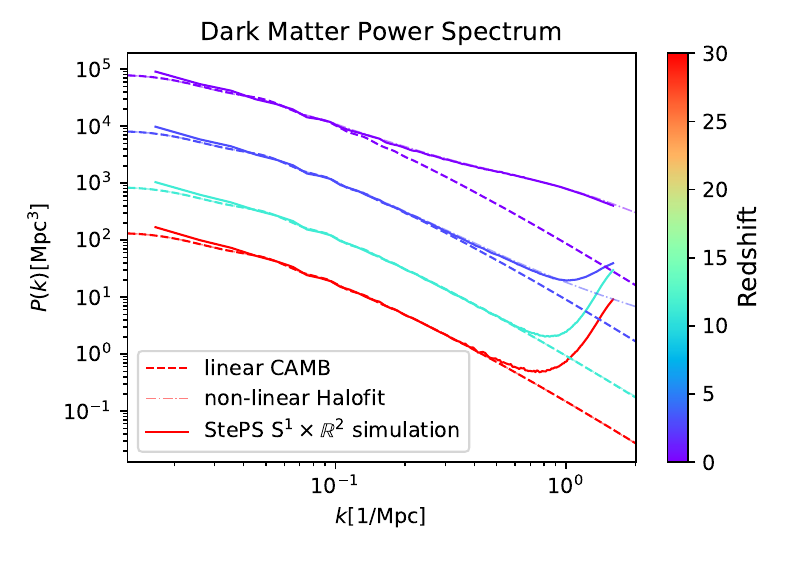}
    \caption{Dark matter power spectra of four selected snapshots (solid lines) with the corresponding theoretical linear power spectra (dashed lines) and the non-linear \cite{2021MNRAS.502.1401M} halofit power spectra (dotted dashed lines). Our \SoneRtwo \StePS simulation is capable of following both linear and non-linear structure formation.}
    \label{fig:PowerSpectra}
\end{figure}

The isotropic $P(k)$ power spectrum of the dark matter density field is a standard statistic used to quantify the clustering. This is defined as
\begin{equation}
\left\langle \delta(\vec{k})\,\delta^*(\vec{k}') \right\rangle
= (2\pi)^3\,\delta_{\rm D}(\vec{k}-\vec{k}')\,P(k),
\qquad k\equiv|\vec{k}|,
\label{eq:Pk_continuous_def}
\end{equation}
where $\delta_{\rm D}$ is the Dirac delta function and the angle brackets denote an ensemble average and 
\begin{equation}
\delta(\vec{k}) = \int \delta(\vec{x})\,e^{-i\vec{k}\cdot\vec{x}}\,{\rm d}^3x = \int \frac{\rho_{\rm DM}(\vec{x})-\bar{\rho}_{\rm DM}}{\bar{\rho}_{\rm DM}} \,e^{-i\vec{k}\cdot\vec{x}}\,{\rm d}^3x
\label{eq:delta_k_continuous}
\end{equation}
is the Fourier transform of the dark matter density contrast of the simulation. In a periodic \Tcube volume, $P(k)$ can be estimated efficiently using FFTs on a uniform grid. In the \SoneRtwo\ topology, however, the \StePS zoom-in particle load implies a strongly position-dependent effective sampling and a non-trivial window, so making a direct FFT-based estimate on a single uniform grid is not straightforward. We therefore estimated the power spectra from particle snapshots using the Feldman--Kaiser--Peacock (FKP) method \citep{1994ApJ...426...23F}. The FKP estimator was originally developed to measure galaxy power spectra from redshift surveys with non-uniform selection functions and complex survey geometries. It constructs a weighted fluctuation field by subtracting a synthetic, unclustered `random' catalogue that follows the same selection function and window as the data and then it applies a Fourier transform to this weighted field. This random catalogue in our case is the initial conditions that was used during glass making.
The estimated power spectra of a few selected snapshots can be seen in Fig.~\ref{fig:PowerSpectra}. The estimated power spectra follow the back-scaled \CAMB predictions on linear scales, while showing a good agreement with the non-linear \cite{2021MNRAS.502.1401M} halofit power spectrum emulator,
which was calibrated on large cubic \Tcube simulations.

\subsection{Dark matter halos}

The dark matter halos are gravitationally bound dark matter regions that have decoupled from the cosmic expansion and collapsed \citep{2018ARA&A..56..435W}. While these halos are the prospective hosts of galaxies, the present runs are dark-matter-only, so individual galaxies are not explicitly resolved in our simulations. As a consequence of the \SoneRtwo zoom-in configuration, low-mass halos and the fine internal structure of halos can be captured with high fidelity near the axis, while more massive halos are resolvable within larger volumes, according to the $M(\varrho)$ mass resolution profile in Fig.~\ref{fig:S1R2StePS_resolution}. As existing halo finders are not designed for \SoneRtwo zoom-in geometries, we implemented our own spherical-overdensity (SO) halo finder, \StePSHF (R\'acz et al. in prep.).

\StePSHF follows the SO algorithm \citep{1998MNRAS.299..728T, 2016MNRAS.456.2486D} widely used in cosmological analyses, adapted to the $\mathrm{S}^1$ periodicity along the $z$ direction and the non‑compact, radially varying the resolution in the transverse plane. After loading one snapshot, we reconstructed a local density for each particle using a $k$‑nearest‑neighbour estimator that is robust to inhomogeneous sampling. We then identified the highest‑density unflagged particle as a provisional centre and grew a sphere around it, accumulating mass until the mean enclosed density first dropped below the chosen threshold, $\bar{\rho}(<R_{\rm vir}) = \rho_{\rm vir}(z)$, where $\rho_{\rm vir}$ is the redshift‑dependent virial density from spherical collapse \citep{1998ApJ...495...80B}. Whenever the resulting candidate contained at least $N_{\rm p,min}$ particles, we refined its centre by replacing the seed position with the centre of mass of the innermost $N_{\rm p,min}$ particles and recomputing essential parameters $(R_{\rm vir},M_{\rm vir})$, other mass definitions ($M_{200c}$, $M_{200b}$, $M_{500c}$, $M_{1000c}$, etc.), the corresponding radii, and other halo properties such as the peculiar velocity ($\vec{V}$), Klypin scale radius ($R_{\rm Klypin}$, \citealt{2011ApJ...740..102K}), concentration ($c$), angular momentum ($\vec{J}$), Bullock-spin ($\lambda_B$, \citealt{2001ApJ...555..240B}), and total energy ($E$). All particles within $R_{\rm vir}$ were then flagged as members and the process was repeated on the remaining unflagged particles until no seeds above threshold remain.

\begin{figure*}[h!]
    \centering
    \begin{subfigure}[a]{0.3225\textwidth}
        \centering
        \includegraphics[width=\linewidth]{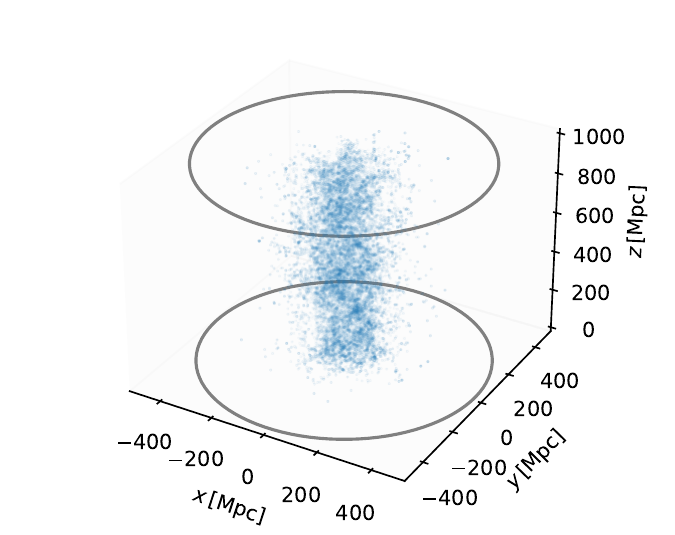}
        \caption{$z=2.0$, $N_{\rm halos} = 3.3\cdot10^{4}$}
    \end{subfigure}
    ~
    \begin{subfigure}[a]{0.3225\textwidth}
        \centering
        \includegraphics[width=\linewidth]{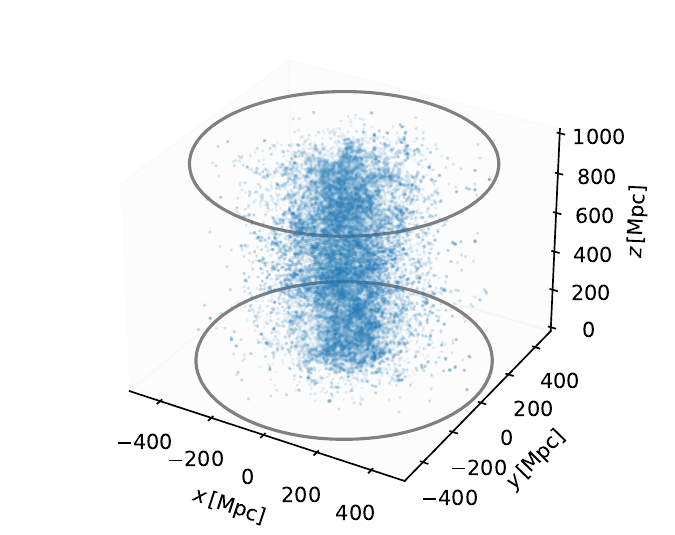}
        \caption{$z=1.0$, $N_{\rm halos} = 7.5\cdot10^{4}$}
    \end{subfigure}
    ~
    \begin{subfigure}[a]{0.3225\textwidth}
        \centering
        \includegraphics[width=\linewidth]{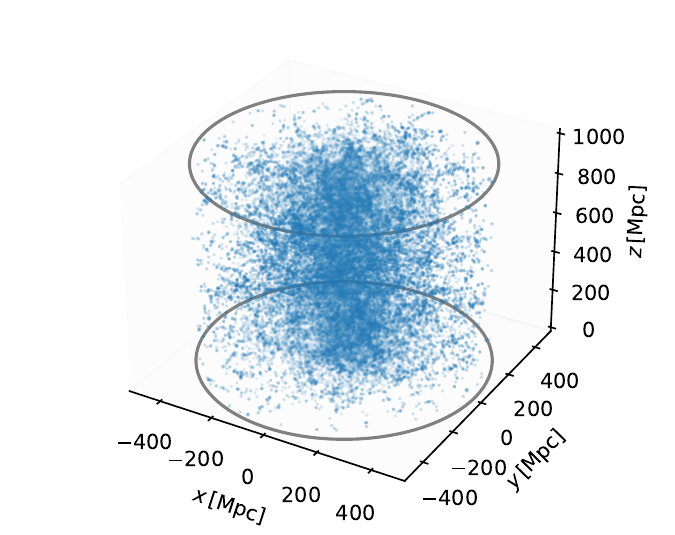}
        \caption{$z=0.0$, $N_{\rm halos} = 1.1\cdot10^{5}$}
    \end{subfigure}
    \caption{Spatial distribution of dark matter halos in the new \SoneRtwo \StePS simulation at different redshifts, in comoving coordinates. Due to the zoom-in geometry, and the fact that low-mass halos form first, halos first appear close to the central axis, where the resolution is highest, and the smallest halos can be resolved. As structure formation proceeds towards higher masses,  halos can later be found at larger radii, where the resolution is lower, and only more massive halos can be resolved. Towards larger $\varrho=\sqrt{x^2+y^2}$ only increasingly massive halos are resolved, so the apparent decline in number density at large $\varrho$ is a selection effect.}
    \label{fig:HaloDist}
\end{figure*}

We used $N_{\rm p,min}=15$ as the minimum threshold for halo particles and identified dark matter halos in late-time snapshots. While this minimal particle number is usually sufficient to identify halo positions, a higher threshold is needed for the reliable measurement of structural parameters \citep{2011MNRAS.415.2293K}. At the final state (redshift 0), we found $1.1\cdot10^{5}$ dark matter halos. The spatial distribution of these halos is visualised in Fig~\ref{fig:HaloDist} at three selected redshifts. Overall, 68 parameters were calculated for each identified halo.

\subsection{Identifying signals of anisotropy}

\begin{figure*}[h!]
    \centering
    \begin{subfigure}[a]{0.3225\textwidth}
        \centering
        \includegraphics[width=\linewidth]{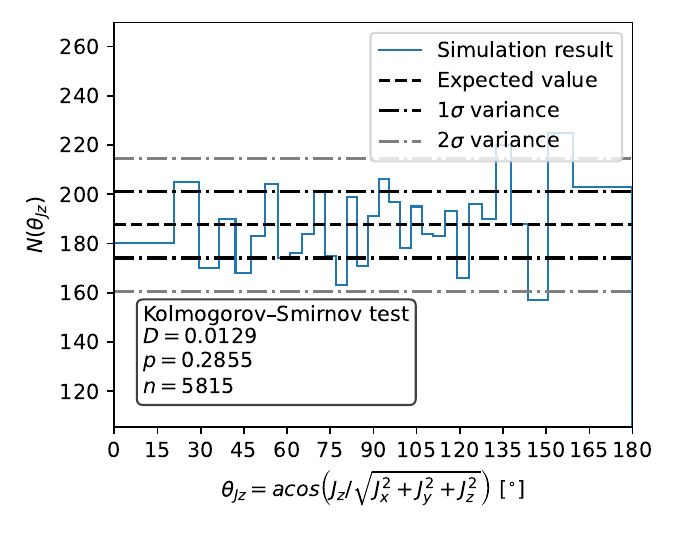}
        \caption{$1.0\cdot10^{13}\msol < M_{vir} < 5.0\cdot10^{13}\msol$}
    \end{subfigure}
    ~
    \begin{subfigure}[a]{0.3225\textwidth}
        \centering
        \includegraphics[width=\linewidth]{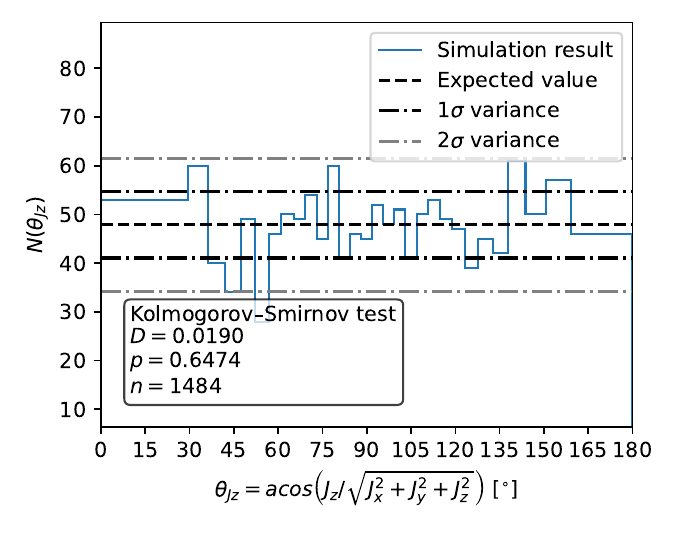}
        \caption{$5.0\cdot10^{13}\msol < M_{vir} < 1.0\cdot 10^{14}\msol$}
    \end{subfigure}
    ~
    \begin{subfigure}[a]{0.3225\textwidth}
        \centering
        \includegraphics[width=\linewidth]{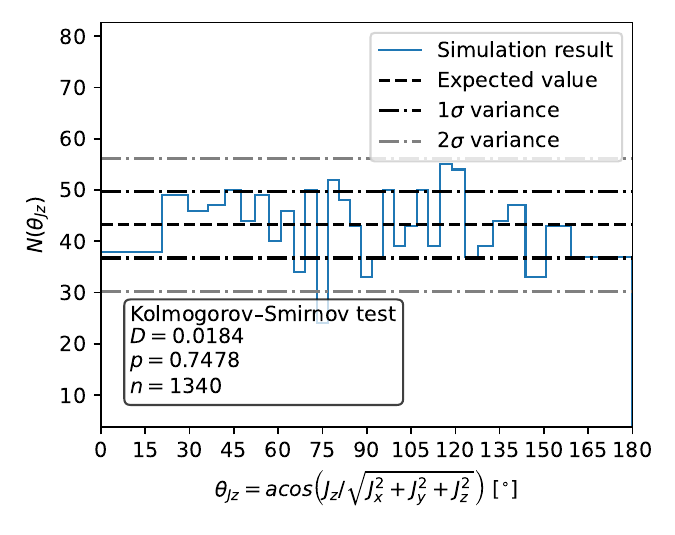}
        \caption{$1.0\cdot 10^{14}\msol < M_{vir} < 1.0\cdot10^{16}\msol$}
    \end{subfigure}
    \caption{Distribution of the halo spin orientation angle, $\theta_{Jz}$, with respect to the simulation $z$-axis, shown for three mass bins at redshift $z=0.5$ in our \SoneRtwo \lcdm simulation. We only used halos with $N_{\rm part}>100$ in this analysis to reliably measure the angular momentum vectors. The histogram uses equal‑area binning on the unit sphere, so that an isotropic spin distribution implies equal expected counts per bin. Each panel is annotated with the one‑sample Kolmogorov-Smirnov test on $u=\frac{1}{2}(1+\cos(\theta_{Jz}))\,\in\,[0,1]$ against the uniform null. The three mass bins show no departures from the isotropic expectation. We found similar distributions in all other output redshifts and mass bins with no clear signs of anisotropy.}
    \label{fig:HaloSpinDist}
\end{figure*}

A recent analysis of the James Webb Space Telescope (JWST) Advanced Deep Extragalactic Survey (JADES)  reported a statistically significant deviation from isotropy in galaxy spin directions \citep{2025MNRAS.538...76S}. Motivated by these findings and by the fact that the gravity is inherently anisotropic in \SoneRtwo topology, we have aimed to detect corresponding late-time signatures of anisotropy in our simulation. In our setup, the compact direction is the periodic $z$-axis and the manifold retains an $\mathrm{SO}(2)$ rotational symmetry around this axis. Since we assumed statistical isotropy during the initial condition generation, any global anisotropy at the halo scales should appear as a deviation from uniformity in the distribution of the angle
\begin{equation}
    \theta_{Jz} = \cos^{-1}\left(\frac{J_{z}}{|\vec{J}|}\right)
\end{equation} between the halo angular-momentum vector, $\vec{J}$, and the $z$-axis. We measured $\vec{J}$ for every halo with $N_{\rm part}\geq 100$ particles to ensure accurate angular momenta, binned the halos by mass and tested the distribution against the isotropic expectation and plotted the results in Fig.~\ref{fig:HaloSpinDist}. Isotropy was quantified by testing uniformity of $\cos\theta_{Jz}$ with Kolmogorov–Smirnov tests in each mass bin. Our result shows that for $L_z=1.0\,\gpc$ comoving periodicity, within our statistical precision and for all mass bins considered, the halo spin–axis distribution is consistent with statistical isotropy in our \lcdm simulation.
We emphasise that smaller compact lengths could in principle enhance anisotropy, but such values are strongly constrained by CMB observations \citep{2016A&A...594A..18P} and by modern large-scale galaxy surveys \citep{1997MNRAS.292..105R, 2002MNRAS.336L..13H, 2016Univ....2....1L}. Our result shows that the \SoneRtwo topology, at least for the comoving $L_z=1.0\,\gpc$ periodicity, does not provide a natural explanation for the JWST rotation‑sense asymmetry reported by \citet{2025MNRAS.538...76S}.

\section{Summary}
\label{sec:summary}

We present a new capability of the \StePS simulation code that enables compactified cosmological \Nbody simulations in \SoneRtwo topology. We summarise the large-scale modifications of the gravitational field induced by this slab periodic boundary condition, described the simulation algorithm, and outline the initial condition generation procedure. We present the first \SoneRtwo \lcdm simulation run with our \StePS code and show that it reproduces the expected linear and non-linear growth of structures. The \lcdm simulation particle outputs, post-processed data, and example notebooks are freely available on our project webpage\footnote{\url{https://eltevo.github.io/projects/cylindrical-simulations}}.

The new simulation topology offers several practical advantages compared to the usual \Tcube setup:
\begin{itemize}
    \item Since the angular momentum in this topology is conserved along the axis of periodicity, the \SoneRtwo \StePS simulations are ideal for studying systems where this quantity plays an important role; for instance, in the context of cosmic filaments and their interactions with dark matter halos or anisotropic cosmological models, such as Bianchi type VII models.
    \item The zoom-in implementation improves the balance between small- and large-scale modes relative to a constant-resolution $\mathrm{T}^3$ run: small scales are resolved at high fidelity only within $\omega<\omega_c$ around the central axis, while long-wavelength modes are sampled over substantially larger volumes.
    \item Similarly, the halo catalogues sample the smallest halos near the axis, while more massive and rarer halos are sampled from much larger volumes due to the zoom-in strategy in $\mathbb{R}^2$.
\end{itemize}
These benefits come with some limitations:
\begin{itemize}
\item Since almost all simulation post-processing pipelines assume \Tcube geometry, processing the outputs of \StePS usually requires heavy modifications in existing pipelines.
\item The zoom-in strategy introduces selection effects and sampling challenges that are absent in constant-resolution \Tcube simulations and these must be accounted for.
\item The anisotropic force field induced by the topology can imprint artefacts, similarly to known issues in the \Tcube case \citep{2021MNRAS.503.5638R}.  Based on preliminary tests, such effects might become non-negligible at late times if the periodic length is small, $L_z\,\lesssim\,200\,\mpc$. A detailed characterisation of these systematics will be presented in future work.
\end{itemize}
We do not expect that the new algorithm will replace the well-tested cubic \Tcube simulations; rather, the aim of this work is to provide an alternative for cases where cubic symmetry can limit the results. This new simulation method is now part of the version 2 release of \StePS along  the \Rcube and \Tcube topological manifolds. This simulation code is freely available as an open-source project at our \textit{github}\footnote{\url{https://github.com/eltevo/StePS/}} repository under GNU General Public License v2.0.

\begin{acknowledgements}
GR and TS acknowledge support of the Research Council of Finland grant 354905 and support by the European Research Council via ERC Consolidator grant KETJU (no. 818930).
This work has been supported by the Hungarian National Research, Development and Innovation Office (NKFIH, grant OTKA NN147550).
IS acknowledges NASA ROSES grants 80NSSC24K1489 and 24-ADAP24-0074, contract number 80NM0018F0610 via a JPL sub-award.
The authors wish to acknowledge CSC – IT Center for Science, Finland, for computational resources.
\end{acknowledgements}

\bibliographystyle{aa}
\bibliography{References}

\end{document}